\documentclass[groupedaddress,aps,pra, reprint]{revtex4-2}

\usepackage{graphicx}
\usepackage{multirow}
\usepackage{amsmath,amssymb,amsfonts}
\usepackage{amsthm}
\usepackage{mathrsfs}

\usepackage{hyperref}
\usepackage{scholax}
\usepackage[scaled=1.075,ncf,vvarbb]{newtxmath}
\usepackage{mathrsfs}
\usepackage{siunitx}
\usepackage{gensymb}
\usepackage{setspace}
\newcommand{\mrm}{\mathrm}
\usepackage{todonotes}
\setuptodonotes{inline}

\newcommand{\Bsca}{\mathcal{B}}

\begin{document}

\title{Enhanced coherence of rare-earth nuclear spins in a crystal}

\author{Bassam Nima}
\author{Yuiki Takahashi}
\author{Amar Vutha}
\affiliation{Department of Physics, University of Toronto, Toronto, Ontario M5S 1A7, Canada}

\begin{abstract}
Rare-earth ions in crystals are useful in precision measurements and quantum information science because of their exceptional optical and spin coherence properties. We show that the nuclear spin coherence time of $^{153}\mathrm{Eu}^{3+}$ ions doped into yttrium orthosilicate can be significantly enhanced by driving the environmental nuclear spins in the host crystal. This method of enhancing the spin coherence time can lead to improved precision measurements using rare-earth-doped crystals.
\end{abstract}
\maketitle

Rare-earth ions doped into yttrium orthosilicate (YSO) crystals are a promising and versatile platform for quantum information science because of their outstanding optical and spin coherence characteristics~\cite{guo2023rare}.
Such systems have been widely applied in quantum memories, quantum transducers and quantum networking. Recently, precision measurements of spin resonances in $^{153}$Eu-doped YSO have been used to probe nuclear time-reversal (T) symmetry violation, demonstrating that rare-earth-doped solids can be used to search for new physics at energy scales beyond 10 TeV \cite{Nima2026} and aid in the hunt for dark matter \cite{Fan2026}. The energetic reach of such measurements can be expanded by improving the precision of nuclear spin resonance measurements. 

The precision of a spin resonance frequency measurement using the Ramsey method \cite{Ramsey1949} is $\delta \nu \propto 1/T_R$ when $T_R \ll t_\mathrm{cycle}$, where $T_R$ is the time interval between time-separated spectroscopy pulses and $t_\mathrm{cycle}$ is the measurement duty cycle. The upper limit on $T_R$ is effectively set by the coherence time, $T_2$, of the spin system. The coherence time of $^{153}$Eu nuclei in YSO is $T_2 \sim 25$ ms, limited by magnetic field noise from the bath of $^{89}$Y ($I=1/2$) and $^{29}$Si ($I=1/2$) nuclear spins in the crystal \cite{Pignol2024}.

Extremely long coherence times in Eu:YSO have been reported using the zero-first-order-Zeeman (ZEFOZ) technique~\cite{zhong_optically_2015, rancic_coherence_2018, Wang2025}: in this technique a pair of hyperfine states of $^{151}$Eu$^{3+}$ ions are tuned, using a magnetic field oriented in a particular direction relative to the crystal axes, to an operating point where the nuclear spin expectation values in the two states are identical. However, the two nuclear spin states also have zero sensitivity to T-violating new physics when tuned to a ZEFOZ point \cite{Radak2026}, which makes this technique unsuitable for extending the precision of nuclear T-violation searches. An alternative technique for improving the coherence time is dynamical decoupling, using a series of $\pi$ pulses applied to the nuclear spin, which has been used to demonstrate coherence time extension to $\sim 0.5$ s in $^{151}$Eu:YSO \cite{Arcangeli2014}. Unfortunately, dynamical decoupling is also not compatible with the Ramsey spectroscopy technique used in precision measurements of T-violation. 

We demonstrate enhancement of the coherence time of $^{153}$Eu nuclear spins in Eu:YSO by resonantly driving the environmental nuclear spins to average away their interaction with $^{153}$Eu nuclei. We find the coherence time can be increased by a factor of $\sim 7$ without compromising the T-violation sensitivity of the nuclear spin states. Importantly, this method of coherence time enhancement can be used during Ramsey spectroscopy measurements of T-violating frequency shifts in $^{153}$Eu:YSO. 
 
Similar work to extend spin coherence dates back to the early days of magnetic resonance in crystals \cite{Sarles1958,*Bloch1958}, but to our knowledge, such a method has not been demonstrated in rare-earth-doped crystals. Beyond the utility of this method in precision measurements such as Refs.\ \cite{Nima2026,Fan2026}, it could offer useful improvements to experiments using rare-earth ions as quantum memories~\cite{Hedges2010} or quantum transducers~\cite{FernandezGonzalvo2015, Rochman2023}. 
\begin{figure*}[hbt!]
    \centering
    \includegraphics[width=\textwidth]{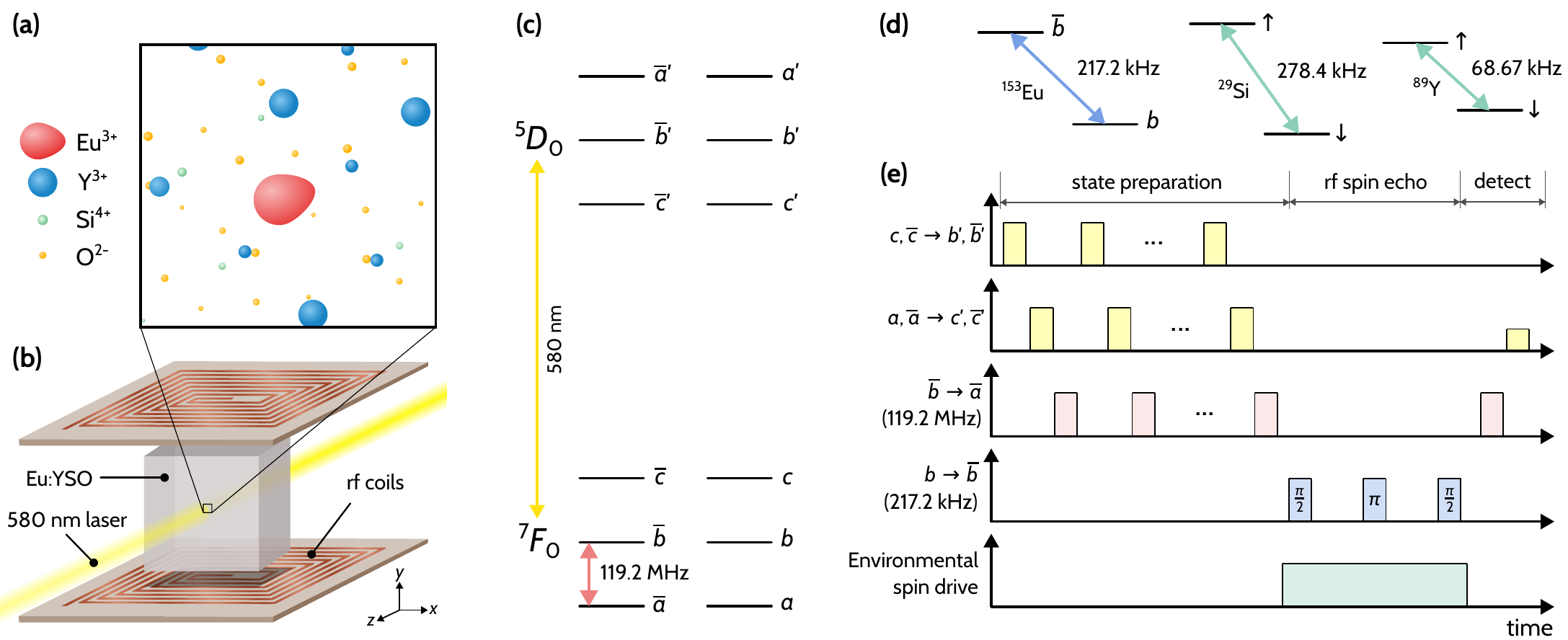}
    \caption{(a) Illustration of a Y$_{2}$SiO$_5$ unit cell in the $x-y$ plane, centered on a ${}^{153}$Eu$^{3+}$ ion doped at site 1. (b) Schematic of the experimental apparatus, showing the Eu:YSO crystal, probe laser and planar rf coils. (c) Energy levels of $^{153}$Eu$^{3+}$ in YSO. The electronic ground ${}^7F_0$ and excited ${}^5D_0$ states are connected by the 580 nm optical transition used for state-preparation and detection. The nuclear spin sublevels are labeled $\bar{a},a; \bar{b},b; \bar{c},c$ in the $^7F_0$ electronic state, and $\bar{a}',a'; \bar{b}',b'; \bar{c}',c'$ in the $^5D_0$ electronic state. (d) The $b-\bar{b}$ transition of $^{153}$Eu and the transitions of environment spins, $ ^{29}$Si and $^{89}$Y, in a $\mathcal{B}\approx340$ G magnetic field. (e) The spin echo sequence used in the experiments. Initial state preparation was done using repeated $c,\bar{c}\rightarrow b', \bar{b}'$, followed by $a,\bar{a}\rightarrow c', \bar{c}'$, and an adiabatic $\bar{b}\rightarrow \bar{a}$ rf sweep. State readout was performed with a $\bar{b}\rightarrow \bar{a}$ rf sweep, followed by $a,\bar{a}\rightarrow c', \bar{c}'$ optical absorption detection. The $b-\bar{b}$ spin echo was measured with/without the $^{89}$Y, $^{29}$Si environmental spin drives (ESD) to measure the coherence enhancement shown in Fig.\ \ref{fig:coherence_enhancement}. }
    \label{fig:cartoon}
\end{figure*}

\textit{Measurements --} The measurements reported here used a 3.5 mm ($D1$) $\times$ 4.0 mm ($D2$) $\times$ 1.0 mm ($b$) YSO crystal doped with 0.1\% Eu, held between samarium-cobalt magnets in a cryostat at $3.5$~K. We label the crystal's dielectric axes $(D1, D2, b)$ as $(x,y,z)$. The experimental apparatus, shown in Fig. \ref{fig:cartoon}b, was similar to that described in Refs.~\cite{Nima2026,Fan2026}.

The energy level structure of $^{153}$Eu$^{3+}$ in YSO is illustrated in Fig. \ref{fig:cartoon}c. A 580 nm optical transition connects the electronic ground state $^7F_0$ with the electronic excited state $^5D_0$. Within the crystal, the quadrupole interaction splits the 6 nuclear sublevels of $^{153}$Eu into 3 Kramers-degenerate pairs in the absence of a magnetic field. We label these pairs as $\bar{a},a; \bar{b},b; \bar{c},c$ in the $^7F_0$ electronic state, and $\bar{a}',a'; \bar{b}',b'; \bar{c}',c'$ in the $^5D_0$ electronic state. The magnetic field applied to the crystal lifts the degeneracy between the Kramers pairs.

A laser was used for nuclear spin state preparation and detection along $z$, and two radio-frequency (rf) coils were used for spectroscopy and spin echo measurements. One coil oriented along $z$ was tuned near $119.2~\mathrm{MHz}$, while the other oriented along $y$ was driven in a lower frequency band ($<1$ MHz), in order to address different hyperfine transitions between the nuclear spin states. The laser used for state-preparation and detection does not distinguish between the states within a Kramers pair, as its spectral linewidth is larger than the separation between the states. The nuclear spin was initialized into the $^{7}F_{0},b$ state, using optical pumping on the $^{7}F_{0},c,\bar{c} \rightarrow {}^{5}D_{0},b',\bar{b}'$ and $^{7}F_{0},a,\bar{a} \rightarrow {}^{5}D_{0},c',\bar{c}'$ transitions, followed by an adiabatic rf sweep centered near $119~\mathrm{MHz}$ to swap population between $^{7}F_{0},\bar{b}$ to $^{7}F_{0},\bar{a}$. This cycle was repeated six times to achieve a population imbalance between $^{7}F_{0},b$ and $^{7}F_{0},\bar{b}$.

The T-violation-sensitive $^{7}F_{0},b \rightarrow {}^{7}F_{0},\bar{b}$ resonance is at $f_{b \bar{b}}$ = 217.2 kHz in the $\Bsca_x$ = 340 G magnetic field used in this experiment. A Hahn echo sequence \cite{Hahn1950} was used to measure the coherence time, $T_2$, of this transition, as shown in Fig.\ \ref{fig:cartoon}e. The echo sequence  consisted of an initial $\pi/2$ pulse, followed by free evolution for a time $\tau$, a $\pi$ pulse (with a phase delay of $90^\circ$ compared to the $\pi/2$ pulses), a second free-evolution step for time $\tau$, and a final $\pi/2$ pulse. We will refer to $\tau$ as the echo time.

\begin{figure}[hbt!]
    \centering
    \includegraphics[trim=0 0.7cm 0 0, clip, width=0.48\textwidth]{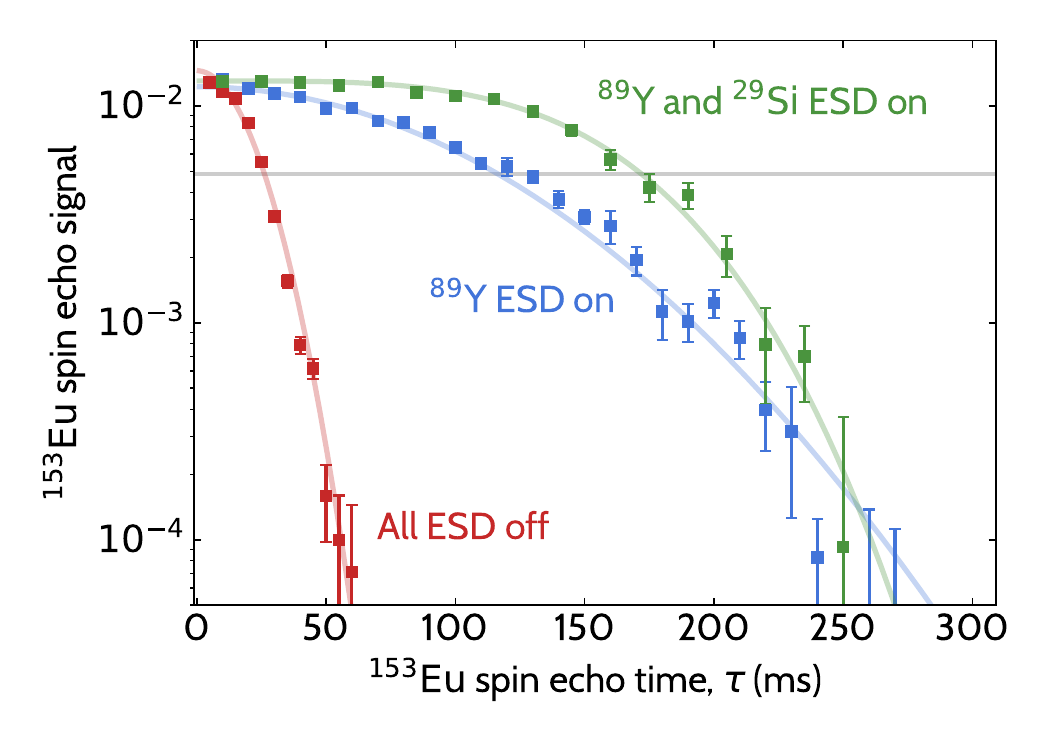}
    \caption{Spin echo decays of the $b-\bar{b}$ transition in ${}^{153}$Eu. The spin echo signal plotted on the vertical axis is in units of optical depth. The fits show the unperturbed $T_2 = 25.0 (2)$ ms (red) increasing to $T_2' = 122 (1)$ ms (blue) with the $^{89}$Y environmental spin drive (ESD) and $T_2'' = 173 (2)$ ms (green) with both $^{89}$Y and $^{29}$Si ESD. The horizontal line indicates a factor of $1/e$ reduction in the spin echo signal.}
    \label{fig:coherence_enhancement}
\end{figure}

\begin{figure*}[hbt!]
    \centering
    \includegraphics[trim=0 0.35cm 0 0, clip, width=\textwidth]{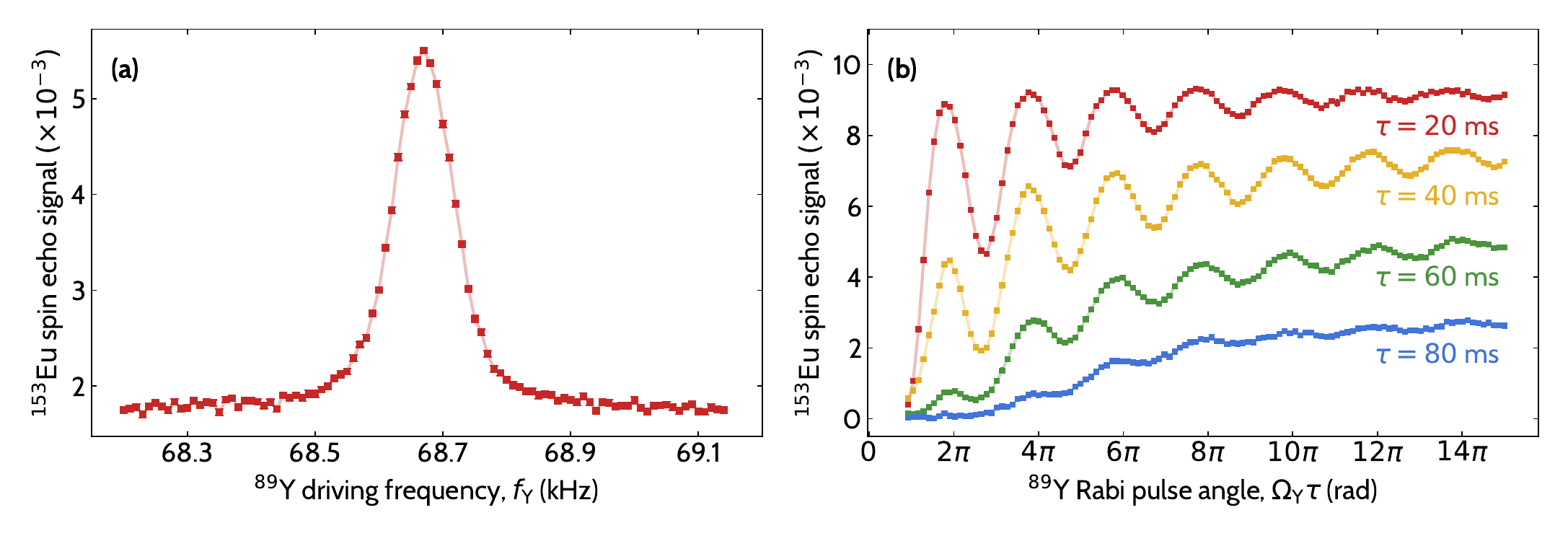}
    \caption{(a) Nuclear spin resonance in ${}^{89}$Y$^{3+}$ measured using spin echoes of $^{153}$Eu$^{3+}$. Fixing the echo time to $\tau=$ 30 ms, the ${}^{89}$Y driving frequency was scanned to map out this resonance. The line center, $f_\mathrm{Y} \approx 68.67$ kHz, is consistent with the resonance frequency expected for $^{89}$Y in $\mathcal{B} \approx 340$ G. 
    (b) Spin echo signals for the $^{153}$Eu $b-\bar{b}$ transition at different values of $\tau$, as a function of the driving magnetic field amplitude -- and thus the Rabi frequency, $\Omega_\mathrm{Y}$ -- for the ${}^{89}$Y spin transition. The x-axis of the plot is the pulse angle $\Omega_\mathrm{Y} \tau$.}
    \label{fig:Y_resonance}
\end{figure*}

After this spin echo sequence, the population between $^{7}F_{0},\bar{b}$ and $^{7}F_{0},\bar{a}$ was inverted using a final rf sweep. The population in $^{7}F_{0},\bar{a},a$ was measured using the optical depth of absorption of the laser tuned to the $^{7}F_{0},a,\bar{a} \rightarrow {}^{5}D_{0},c',\bar{c}'$ transition. In this way, the sequence of rf and optical pulses measures the relative population change between $^{7}F_{0},b$ and $^{7}F_{0},\bar{b}$. 

We note that the symmetries of the YSO crystal give rise to two types of sub-ensembles, labeled $\rho = \pm 1$ and $\sigma = \pm 1$, that are related by inversion and $xy$-plane-reflection respectively \cite{Nima2026}. No electric field was applied to the crystal during these experiments, and so the measurements used both the $\rho = \pm 1$ sub-ensembles. An applied magnetic field along $z$ was used to separate the $b-\bar{b}$ resonance frequencies for $\sigma = \pm 1$, allowing us to selectively probe the $\sigma = +1$ or $\sigma = -1$ sub-ensembles separately. We observed no differences in the measured coherence time or the environmental spin resonance frequency for $\sigma = \pm1$. In the results reported here, the $\sigma=\pm1$ sub-ensembles were overlapped by setting $\mathcal{B}_z = 0$~G.

\begin{figure*}[hbt!]
    \centering
    \includegraphics[trim=0 0.35cm 0 0, clip, width=\textwidth]{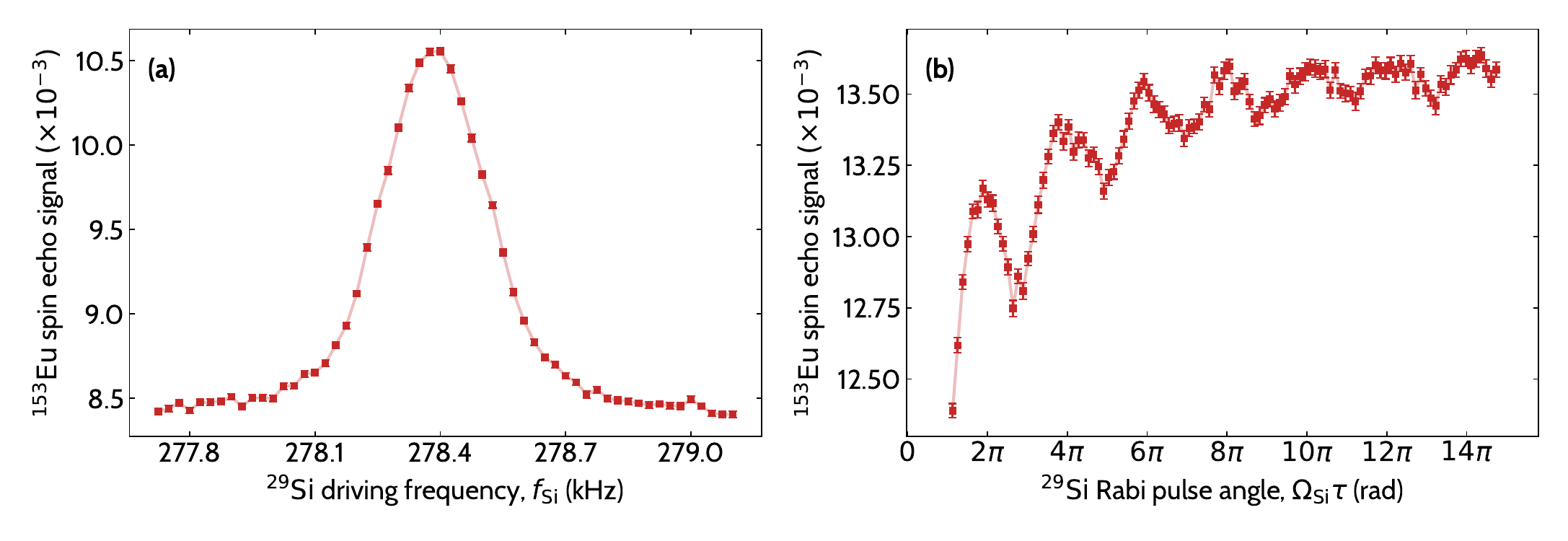}
    \caption{(a) Nuclear spin resonance in ${}^{29}$Si$^{4+}$ measured using spin echoes of $^{153}$Eu$^{3+}$. Fixing the echo time to $\tau=$ 80 ms, the ${}^{29}$Si driving frequency was scanned to map out this resonance. The line center, $f_\mathrm{Si} \approx 278.4$ kHz, is consistent with the resonance frequency expected for $^{29}$Si in $\mathcal{B} \approx 340$ G. 
    (b) Spin echo signal on the $^{153}$Eu $b-\bar{b}$ transition at $\tau = 20$ ms, varying the driving magnetic field amplitude and the Rabi frequency for the ${}^{29}$Si spin transition. The x-axis of the plot is the pulse angle $\Omega_\mathrm{Si} \tau$. For both (a) and (b) the ESD included a $^{89}$Y drive fixed at $f_\mathrm{Y}=68.67$ kHz, $\Omega_\mathrm{Y} = 2\pi \times 1$ kHz. The $^{29}$Si transition was undetectable without applying the $^{89}$Y drive.}
    \label{fig:Si_resonance}
\end{figure*}

The measured spin echo signals as a function of echo time were fitted to $S(\tau) = S_0 \, \exp\left[{-(\tau/T_2)^n} \right]$. The nuclear spin coherence time of $^{153}$Eu without driving the environmental spins fit well to a Gaussian decay profile ($n=2$) with $T_2 = 25.0(2)$~ms, shown in Fig.~\ref{fig:coherence_enhancement}. This value is consistent with the reported coherence times for $^{151}$Eu limited by noise from environmental spins \cite{Alexander2007,Arcangeli2014}. To enhance this coherence time, we applied an environmental spin drive (ESD), consisting of an oscillating rf magnetic field at the environmental spin resonance frequency. The ESD causes the environment spins to switch orientations rapidly, averaging away their interactions with the $^{153}$Eu spins of interest. When we applied a single tone ESD at the resonance frequency of $^{89}$Y, we observed a large enhancement of the $^{153}$Eu spin echo signal. As shown in Fig. \ref{fig:coherence_enhancement}, the coherence time increased to  $T_2' = 122(1)$ ms while retaining its Gaussian decay profile ($n=2$). We then applied two ESD tones, resonant with $^{89}$Y and $^{29}$Si, which increased the coherence time further, to  $T_2''=173(2)$ ms with $n =3.8(2)$. We observed no enhancement with only $^{29}$Si ESD, suggesting that $^{89}$Y nuclear spins are the dominant contribution to the dephasing of $^{153}$Eu nuclei.

The coherence enhancement effect allows the $^{89}$Y spin resonance to be mapped out as a function of ESD frequency, through measuring the $^{153}$Eu spin echo signal (Fig. \ref{fig:Y_resonance}a). We also observed distinct Rabi oscillations of $^{89}$Y while varying the $^{89}$Y drive amplitude, again through the effect on the $^{153}$Eu spin echo signal. As shown in Fig. \ref{fig:Y_resonance}b, the $^{153}$Eu spin echo signal varies consistently with the pulse angle $\theta_\mathrm{Y} = \Omega_\mathrm{Y} \tau$ as expected for Rabi oscillations. The Rabi frequency $\Omega_\mathrm{Y}$ estimated from these oscillations matches the value calculated from the magnetic field amplitude applied through the drive coil (calibrated using measurements of the rf coil impedance and resonance frequency) and the nuclear magnetic dipole moment of $^{89}$Y ($\mu_\mathrm{Y} = 0.137 \, \mu_N$ \cite{Stone2025}). 

In a similar way, the $^{29}$Si spin resonance was mapped out as shown in Fig. \ref{fig:Si_resonance}a. In this case, the ESD included a fixed $^{89}$Y drive at $f_\mathrm{Y}=68.67$ kHz, $\Omega_\mathrm{Y} = 2\pi \times 1$ kHz, in addition to a variable tone $f_\mrm{Si}$. We measured the effect of the $^{29}$Si drive amplitude on the $^{153}$Eu spin echo signal, as a function of the pulse angle $\theta_\mathrm{Si} = \Omega_\mathrm{Si} \tau$, as shown in \ref{fig:Si_resonance}b. Here too, the observed Rabi frequency $\Omega_\mathrm{Si}$ is consistent with the calculated value based on the calibrated magnetic field amplitude and the nuclear magnetic dipole moment of $^{29}$Si ($\mu_\mathrm{Si} = 0.555 \, \mu_N$ \cite{Stone2025}). The decay of the $^{153}$Eu spin echo signal is non-Gaussian when both $^{89}$Y and $^{29}$Si environmental spins are driven (Fig.\ \ref{fig:coherence_enhancement}). The origin of this behaviour and the ultimate limit to the coherence time of $^{153}$Eu in YSO remain to be understood.
~\\~\\
\textit{Implications for T-violation searches --} The enhancement of the coherence time demonstrated here is a step towards improving the sensitivity of precision measurements that use rare-earth ions in YSO, such as searches for the T-violating nuclear Schiff moment \cite{Nima2026} and ultralight dark matter \cite{Fan2026}. In the measurements reported here, the value of $T_R$ was limited by inhomogeneity in the DC magnetic field. Translating the coherence enhancement demonstrated here into improved frequency precision requires magnetic field homogeneity at the level of 20 ppm, which is achievable using standard field-shimming techniques.

In the recent T-violation measurement using $^{153}$Eu:YSO \cite{Nima2026} the Ramsey time was $T_R \approx 1.9$~ms and the total cycle time was  $t_\mathrm{cycle} = 1.58$~s. Since the cycle time for the measurement in \cite{Nima2026} was dominated by state preparation and detection, using the coherence enhancement method reported here to increase $T_R$ to 173 ms would linearly improve the frequency precision, $\delta\nu$, by a factor $173/1.9 \approx 90$, with corresponding improvement in the new physics sensitivity of the measurement. 

However, it is not just the statistical sensitivity that matters in precision measurements. Here we analyze the possibility of potential systematic errors due to AC Zeeman shifts that are introduced by the ESD. Driving the environmental spins during the $^{153}$Eu Ramsey measurement leads to the AC Zeeman shift 
\begin{equation}
    \Delta \omega_{\mathrm{ACZ}} = 
\frac{\omega_\mathrm{Eu}^3}{2}
\left(\frac{(\Omega_\mathrm{Y}/\omega_\mathrm{Y})^2}{\omega_{\mathrm{Eu}}^2 -\omega_\mathrm{Y}^2} 
+ 
\frac{(\Omega_\mathrm{Si}/\omega_\mathrm{Si})^2}{\omega_{\mathrm{Eu}}^2 -\omega_\mathrm{Si}^2}\right),
\end{equation}
where $\omega_\mathrm{Eu}\Omega_{\mathrm{Y}}/\omega_\mathrm{Y}$ and $\omega_\mathrm{Eu}\Omega_{\mathrm{Si}}/\omega_\mathrm{Si}$ are the Rabi frequencies of the transition in Eu, and $\omega_{\mathrm{Eu}}$, $\omega_\mathrm{Y}$ and $\omega_\mathrm{Si}$ are the resonance frequencies of the Eu, Y and Si transitions, respectively. Using $\Omega_{\mathrm{Si}} \approx  \Omega_\mathrm{Y} \approx 2 \pi \times 1$ kHz, 
$\omega_{\mathrm{Eu}} = 2 \pi f_{b \bar{b}} = 2\pi \times 217.2$ kHz, $\omega_\mathrm{Y} = 2 \pi \times 68.67$ kHz and $\omega_{\mathrm{Si}} = 2 \pi \times 278.4$ kHz, we estimate $\Delta \omega_{\mathrm{ACZ}} \approx 2\pi \times 23$ Hz. The magnitude of this frequency shift is large compared to the accuracy achieved in recent precision measurements in Eu:YSO~\cite{Nima2026}. 

Conveniently in $^{153}$Eu:YSO, since the values of $\omega_\mrm{Eu}^2 - \omega_\mrm{Y}^2$ and $\omega_\mrm{Eu}^2 - \omega_\mrm{Si}^2$ have opposite signs, it is possible to cancel the AC Zeeman effect by controlling the ratio of $\Omega_\mrm{Si}$ and $\Omega_\mrm{Y}$. Since only the ratio of Rabi frequencies needs to be fixed for cancelling the AC Zeeman shift, the magnitudes of $\Omega_\mrm{Si}$ and $\Omega_\mrm{Y}$ can be adjusted to be adequately strong for coherence enhancement. The condition for AC Zeeman shift cancellation can be written in terms of the ESD magnetic field amplitudes $\Bsca_\mrm{Y}, \Bsca_\mrm{Si}$ as 
\begin{equation}
    \frac{\Bsca_\mrm{Si}}{\Bsca_\mrm{Y}} = \sqrt{\frac{\omega_\mrm{Si}^2 - \omega_\mathrm{Eu}^2}{\omega_\mathrm{Eu}^2 - \omega_\mrm{Y}^2}} = 0.845 .
\end{equation}
In addition to this mechanism for AC Zeeman shift cancellation, Eu:YSO also has accurate crystal-symmetry comagnetometers \cite{Nima2025}. Cancellation of common-mode shifts for these comagnetometer ensembles to better than a few parts per million has been demonstrated \cite{Nima2026}. The value of the AC Zeeman shift is identical for the $b-\bar{b}$ transition in $\rho=\pm 1$ sub-ensembles of $^{153}$Eu ions, and thus the comagnetometer comparison can be used to suppress any residual AC Zeeman shifts that are not removed by controlling the Y/Si Rabi frequency ratio. Any further residuals, that somehow survive even the comagnetometer comparison, can be isolated using the fact that the coherence enhancement is essentially unaffected for a range of different values of the ESD Rabi frequencies at long $\tau$ (see Fig. \ref{fig:Y_resonance}b), whereas the AC Zeeman shift has a characteristic dependence on $\Omega_\mathrm{Y}$ and $\Omega_\mathrm{Si}$. Thus, systematic effects due to the environmental driving fields can be suppressed using two levels of cancellation and distinguished from genuine T-violation signals.

In summary, we have demonstrated a practical means to increase the coherence time for $^{153}$Eu nuclear spins in YSO by scrambling the effect of environmental spins. The enhanced coherence time in $^{153}$Eu:YSO makes it possible to significantly improve the sensitivity, and extend the high-energy reach, of T-violation and dark matter search experiments. The method demonstrated in this work does not rely on properties specific to $^{153}\mathrm{Eu}$, and so it can be used to enhance coherence times in other yttrium-containing crystals used in quantum science (e.g., \cite{Kindem2018,Liang2020,Zhou2023}). Scrambling environmental spins in this way could benefit a broad range of solid-state experiments, including $^{229}$Th nuclear clocks in crystals \cite{Ooi2026,Huang2026}.
~\\~\\
\textit{Acknowledgments --} We acknowledge helpful discussions with Mingyu Fan, Andrew Jayich, David Patterson, Lilian Childress, Jonathan Weinstein and Joseph Thywissen. This work was supported by NSERC, and by the John Templeton Foundation (Grant No. 63119) and Alfred P. Sloan Foundation (Grant No. G-2023-21045) through the Small-scale Experiments for Fundamental Physics program. B.N. acknowledges support from an NSERC CGRS Doctoral Fellowship. 
~\\~\\
The data that support the findings of this article are publicly available \cite{nima_2026_23702941}.

\bibliography{ce}

@dataset{nima_2026_23702941,
  author       = {Nima, Bassam and
                  Takahashi, Yuiki and
                  Vutha, Amar},
  title        = {Enhanced coherence of rare-earth nuclear spins in
                   a crystal
                  },
  year         = 2026,
  publisher    = {Zenodo},
  doi          = {10.5281/zenodo.23702941},
  url          = {https://doi.org/10.5281/zenodo.23702941},
}

@article{Bloch1958,
  title = {Theory of Line Narrowing by Double-Frequency Irradiation},
  author = {Bloch, F.},
  journal = {Phys. Rev.},
  volume = {111},
  issue = {3},
  pages = {841--853},
  numpages = {0},
  year = {1958},
  doi = {10.1103/PhysRev.111.841}
}

@article{Huang2026,
  title={A nuclear clock synchronized to $^{229}${Th}},
  author={Huang, Beichen and Yan, Gaowei and Xiao, Qi and Bu, Wenhao and Zhao, Chengchun and Zhang, Zhen and Yan, Chao and Chen, Zhi-Ang and Zhang, Peixiong and Penyazkov, Gleb and others},
  journal={Nature},
  volume={658},
  pages={903--908},
  year={2026}
}

@misc{Stone2025,
  author       = {Stone, N.J.},
  title        = {Table of Recommended Nuclear Magnetic Dipole Moments: {Part} {I}, Long-Lived States},
  year         = 2025,
  doi          = {10.61092/iaea.qs1q-27sa}
}

@article{Liang2020,
author = {Peng-Jun Liang and Xiao Liu and Pei-Yun Li and Zong-Quan Zhou and Chuan-Feng Li and Guang-Can Guo},
journal = {J. Opt. Soc. Am. B},
number = {6},
pages = {1653--1658},
publisher = {Optica Publishing Group},
title = {Spectroscopic investigations of ${}^{142}\mathrm{Nd}^{3+}$:{YVO}$_4$ for quantum memory applications},
volume = {37},
year = {2020},
doi = {10.1364/JOSAB.388740}
}

@article{Kindem2018,
  title = {Characterization of $^{171}\mathrm{Yb}^{3+}$:{YVO}$_4$ for photonic quantum technologies},
  author = {Kindem, Jonathan M. and Bartholomew, John G. and Woodburn, Philip J. T. and Zhong, Tian and Craiciu, Ioana and Cone, Rufus L. and Thiel, Charles W. and Faraon, Andrei},
  journal = {Phys. Rev. B},
  volume = {98},
  issue = {2},
  pages = {024404},
  numpages = {10},
  year = {2018},
  doi = {10.1103/PhysRevB.98.024404}
}

@article{Zhou2023,
author = {Zhou, Zong-Quan and Liu, Chao and Li, Chuan-Feng and Guo, Guang-Can and Oblak, Daniel and Lei, Mi and Faraon, Andrei and Mazzera, Margherita and de Riedmatten, Hugues},
title = {Photonic Integrated Quantum Memory in Rare-Earth Doped Solids},
journal = {Laser \& Photonics Reviews},
volume = {17},
number = {10},
pages = {2300257},
doi = {https://doi.org/10.1002/lpor.202300257},
year = {2023}
}

@article{Ooi2026,
  title={Frequency reproducibility of solid-state {Th}-229 nuclear clocks},
  author={Ooi, Tian and Doyle, Jack F and Zhang, Chuankun and Higgins, Jacob S and Ye, Jun and Beeks, Kjeld and Sikorsky, Tomas and Schumm, Thorsten},
  journal={Nature},
  volume={650},
  pages={72},
  year={2026}
}

@article{Ramsey1949,
  title = {A New Molecular Beam Resonance Method},
  author = {Ramsey, Norman F.},
  journal = {Phys. Rev.},
  volume = {76},
  issue = {7},
  pages = {996},
  numpages = {0},
  year = {1949},
  doi = {10.1103/PhysRev.76.996}
}

@article{Sarles1958,
  title = {Double Nuclear Magnetic Resonance and the Dipole Interaction in Solids},
  author = {Sarles, L. R. and Cotts, R. M.},
  journal = {Phys. Rev.},
  volume = {111},
  issue = {3},
  pages = {853--859},
  numpages = {0},
  year = {1958},
  doi = {10.1103/PhysRev.111.853}
}

@article{Alexander2007,
author = {Annabel L. Alexander and Jevon J. Longdell and Matthew J. Sellars},
journal = {J. Opt. Soc. Am. B},
number = {9},
pages = {2479--2482},
publisher = {Optica Publishing Group},
title = {Measurement of the ground-state hyperfine coherence time of ${}^{151}\mathrm{Eu}^{3+}:\mathrm{Y}_{2}\mathrm{SiO}_5$},
volume = {24},
year = {2007},
doi = {10.1364/JOSAB.24.002479}
}

@article{Arcangeli2014,
  title = {Spectroscopy and coherence lifetime extension of hyperfine transitions in ${}^{151}\mathrm{Eu}^{3+}:\mathrm{Y}_2\mathrm{SiO}_5$},
  author = {Arcangeli, Andrea and Lovri\ifmmode \acute{c}\else \'{c}\fi{}, Marko and Tumino, Biagio and Ferrier, Alban and Goldner, Philippe},
  journal = {Phys. Rev. B},
  volume = {89},
  issue = {18},
  pages = {184305},
  numpages = {6},
  year = {2014},
  doi = {10.1103/PhysRevB.89.184305}
}

@article{Hahn1950,
  title = {Spin Echoes},
  author = {Hahn, E. L.},
  journal = {Phys. Rev.},
  volume = {80},
  issue = {4},
  pages = {580--594},
  numpages = {0},
  year = {1950},
  doi = {10.1103/PhysRev.80.580}
}

@article{Fan2026,
  title = {Wideband Search for Axionlike Dark Matter Using Octupolar Nuclei in a Crystal},
  author = {Fan, Mingyu and Nima, Bassam and Radak, Aleksandar and Alonso-\'Alvarez, Gonzalo and Vutha, Amar},
  journal = {Phys. Rev. Lett.},
  volume = {136},
  issue = {12},
  pages = {121802},
  numpages = {6},
  year = {2026},
  doi = {10.1103/dm9j-9pry}
}

@article{Pignol2024,
  title = {Decoherence induced by dipole-dipole couplings between atomic species in rare earth ion doped {Y}$_2${SiO}$_5$},
  author = {Pignol, C. and Ortu, A. and Nicolas, L. and D'Auria, V. and Tanzilli, S. and Chaneli\`ere, T. and Afzelius, M. and Etesse, J.},
  journal = {Phys. Rev. B},
  volume = {110},
  issue = {21},
  pages = {214208},
  numpages = {18},
  year = {2024},
  doi = {10.1103/PhysRevB.110.214208}
}

@article{FernandezGonzalvo2015,
  author  = {Fernandez-Gonzalvo, Xavier and Chen, Yu-Hui and Yin, Chunming and Rogge, Sven and Longdell, Jevon J.},
  title   = {Coherent frequency up-conversion of microwaves to the optical telecommunications band in an {Er:YSO} crystal},
  journal = {Phys. Rev. A},
  volume  = {92},
  pages   = {062313},
  year    = {2015},
  doi     = {10.1103/PhysRevA.92.062313}
}

@article{Rochman2023,
  author  = {Rochman, Jake and Xie, Tian and Bartholomew, John G. and Schwab, K. C. and Faraon, Andrei},
  title   = {Microwave-to-optical transduction with erbium ions coupled to planar photonic and superconducting resonators},
  journal = {Nature Communications},
  volume  = {14},
  pages   = {1153},
  year    = {2023},
  doi     = {10.1038/s41467-023-36799-0}
}

@article{Hedges2010,
  author  = {Hedges, Morgan P. and Longdell, Jevon J. and Li, Yongmin and Sellars, Matthew J.},
  title   = {Efficient quantum memory for light},
  journal = {Nature},
  year    = {2010},
  volume  = {465},
  number  = {7301},
  pages   = {1052--1056},
  doi     = {10.1038/nature09081},
  url     = {https://doi.org/10.1038/nature09081},
  issn    = {1476-4687}
}

@article{zhong_optically_2015,
	title = {Optically addressable nuclear spins in a solid with a six-hour coherence time},
	volume = {517},
	issn = {0028-0836, 1476-4687},
	url = {https://www.nature.com/articles/nature14025},
	doi = {10.1038/nature14025},
	number = {7533},
	urldate = {2024-10-16},
	journal = {Nature},
	author = {Zhong, Manjin and Hedges, Morgan P. and Ahlefeldt, Rose L. and Bartholomew, John G. and Beavan, Sarah E. and Wittig, Sven M. and Longdell, Jevon J. and Sellars, Matthew J.},
	month = jan,
	year = {2015},
	pages = {177--180},
}

@article{rancic_coherence_2018,
	title = {Coherence time of over a second in a telecom-compatible quantum memory storage material},
	volume = {14},
	doi = {10.1038/nphys4254},
	number = {1},
	urldate = {2026-02-03},
	journal = {Nat. Phys.},
	author = {Rančić, Miloš and Hedges, Morgan P. and Ahlefeldt, Rose L. and Sellars, Matthew J.},
	year = {2018},
	pages = {50--54},
}

@misc{Radak2026,
      title={Sensing {T}-violating nuclear moments of paramagnetic ions in crystals}, 
      author={Aleksandar Radak and Mingyu Fan and Bassam Nima and Yuiki Takahashi and Amar Vutha},
      year={2026},
      eprint={2603.24907},
      archivePrefix={arXiv},
      primaryClass={physics.atom-ph}
}

@article{Nima2026,
  title={Limit on the nuclear {Schiff} moment of europium-153},
  author={Nima, Bassam and Fan, Mingyu and Wang, Xubo and Wang, Sen and Zhou, En Fu and Jayich, Andrew M and Yao, Jiang Ming and Cheng, Lan and Vutha, Amar},
  journal={arXiv:2606.12084},
  year={2026}
}

@article{Nima2025,
  title = {Precision comagnetometry for {T}-violation searches in crystals},
  author = {Nima, Bassam and Fan, Mingyu and Radak, Aleksandar and Jayich, Andrew M. and Vutha, Amar},
  journal = {Phys. Rev. A},
  volume = {112},
  issue = {3},
  pages = {L030801},
  numpages = {6},
  year = {2025},
  doi = {10.1103/jykv-rsd1}
}

@article{Wang2025,
  title = {Nuclear Spins in a Solid Exceeding 10-Hour Coherence Times for Ultra-Long-Term Quantum Storage},
  author = {Wang, Fudong and Ren, Miaomiao and Sun, Weiye and Guo, Mucheng and Sellars, Matthew J. and Ahlefeldt, Rose L. and Bartholomew, John G. and Yao, Juan and Liu, Shuping and Zhong, Manjin},
  journal = {PRX Quantum},
  volume = {6},
  issue = {1},
  pages = {010302},
  numpages = {12},
  year = {2025},
  doi = {10.1103/PRXQuantum.6.010302}
}

@article{guo2023rare,
  title={Rare-earth quantum memories: The experimental status quo},
  author={Guo, Mucheng and Liu, Shuping and Sun, Weiye and Ren, Miaomiao and Wang, Fudong and Zhong, Manjin},
  journal={Frontiers of Physics},
  volume={18},
  number={2},
  pages={21303},
  year={2023},
  publisher={Springer}
}

\end{document}